\documentclass[floatfix,aps,pra,showpacs,twocolumn,amsmath,preprintnumbers,nofootinbib]{revtex4} 

\usepackage{amssymb}
\usepackage{graphicx}
\usepackage[caption=false]{subfig}

\def\art{paper }
\def\jrn#1#2#3#4#5#6{#3 \textbf{#4}, #5 (#6)} \def\andd{and } 

\def\scn#1#2{\section{#1}\lb{#2}}
\def\sscn#1#2{\subsection{#1}\lb{#2}}
\def\bfl{\begin{flushleft}}
\def\efl{\end{flushleft}}
\def\bfr{\begin{flushright}}
\def\efr{\end{flushright}}
\def\bc{\begin{center}}
\def\ec{\end{center}}
\def\be{\begin{equation}}
\def\ee{\end{equation}}
\def\ba{\begin{eqnarray}}
\def\ea{\end{eqnarray}}
\def\baa#1{\begin{array}{#1}}
\def\eaa{\end{array}}
\def\bw{\begin{widetext}}
\def\ew{\end{widetext}}

\def\lb#1{\label{#1}}
\def\bit{\begin{itemize}}
\def\eit{\end{itemize}}
\def\bco{}

\def\twomat#1#2#3#4{\begin{pmatrix} #1 & #2 \\ #3 & #4 \end{pmatrix}}

\def\schrod{Schr\"odinger  }

\def\densnnorm{\hat{\Omega}}
\def\rabi{\omega}

\def\mixmat{\hat{M}}
\def\drho{\hat{\Delta}}
\def\frho{\delta\hat{\rho}}

\def\dgammatr{\langle \Gamma \rangle_\Delta}
\def\fpur{\delta {\cal P}}
\def\charmat{\boldsymbol{\Lambda}}
\def\charvec{\mathbf{X}}
\def\pdmat{\mathbf{P}}

\begin{document}

\preprint{\small Eur. Phys. J. D 69 (2015) 253 [arXiv:1505.03408]}

\title{
Non-Hermitian Hamiltonians and stability of pure states
}

\author{Konstantin G. Zloshchastiev}
\affiliation{Institute of Systems Science, 
Durban University of Technology, P.O. Box 1334, Durban 4000, South Africa}

\email{k.g.zloschastiev@gmail.com}

\begin{abstract}
We demonstrate that quantum fluctuations can cause, under certain conditions, the 
dynamical instability of pure states that can result in their evolution
into mixed states.
It is shown that the degree and type of such an instability
are controlled by the environment-induced anti-Hermitian parts of Hamiltonians. 
Using the quantum-statistical approach for
non-Hermitian Hamiltonians and related non-linear master equation, 
we derive the equations that are necessary to study the
stability properties of any model described by a non-Hermitian Hamiltonian.
It turns out that the instability of pure states
is not preassigned in the evolution equation but arises as the emergent
phenomenon in its solutions. 
In order
to illustrate the general formalism and different types of
instability that may occur, we perform the local stability analysis of
some exactly solvable two-state models, which
can be used in the theories of open quantum-optical and spin systems.
\end{abstract}

\date{\footnotesize 17 March 2015}

\pacs{03.65.-w, 05.30.-d, 05.45.-a\\
}

\maketitle

\section{Introduction}

It is well-known that purity is exactly preserved
during unitary evolution driven by Hermitian Hamiltonians.
This property is natural for describing isolated quantum systems but in the case of open
ones it is no longer compulsory \cite{gzbook,bpbook}.
From the viewpoint of the theory of open quantum systems,
the isolated  system is a mere theoretical idealization --
since in the real world all quantum systems are embedded into a
background of some kind \cite{ser06,ser07}.
Even the process of quantum measurement itself 
fits into this framework,
because it involves interaction of the quantum system,
which is being measured, with an external apparatus.
Correspondingly,
once the system is brought into interaction with its environment, 
such as a heat bath, dissipation usually increases its entropy and pure states are converted
into mixed ones \cite{phz93,zur96}.


Recently, the dynamical behaviour of quantum purity and pure states 
has become of considerable research interest
when studied within the framework of the 
non-Hermitian (NH) formalism \cite{sz13,bg12}.
Non-Hermitian Hamiltonians find numerous applications in many areas of physics 
including 
studies of Feshbach resonances and decaying states, quantum transport and scattering by complex potentials, multiphoton ionization, free-electron lasers and optical resonators and waveguides. 
But the biggest area of application is the
theory of open quantum 
systems 
where the anti-Hermitian part 
appears as a result of 
the interaction of systems with their environment 
\cite{nmbook,suu54,kor64,wong67,heg93,bas93,rotter09,reiter,kar14,sz14,sz14cor,ser15,sz15}.
In this \art
we demonstrate the 
profound
connection between NH Hamiltonians 
and dynamical stability of pure states
for the quantum systems of general type.
We show that 
quantum fluctuations can cause 
the instability of pure states whose degree and type are governed
by 
the environment-induced 
anti-Hermitian terms in Hamiltonians.

The contents of this paper are as follows. 
In section \ref{s-nhgen}, we provide a brief account of the density operator approach for NH dynamics and formulate the essence of the stability problem for pure states. 
In section \ref{s-gen}, we derive the general equations which are needed for stability analysis
of pure states for NH-driven systems and discuss their generic features.  
In section \ref{s-loc}, we study the linearized limit of the stability equations, which
can be used for the analysis of local stability.
In section \ref{s-ex}, we consider some two-level systems in order to illustrate the 
general formalism.
Discussions and conclusions are given in section \ref{s-con}.

\scn{Non-Hermitian dynamics}{s-nhgen}

If the Hamiltonian of a quantum system is a non-Hermitian operator, then
it can be decomposed into its Hermitian and anti-Hermitian parts, respectively:
\be\lb{e:nhham}
\hat H = \hat H_+ + \hat H_-
=
\hat H_+  - i \hat\Gamma
,
\ee
where $\hat{H}_\pm = \pm \hat{H}_\pm^{\dag}$,
and $\hat\Gamma = \hat\Gamma^{\dag} $ is usually dubbed the decay operator.
The probability-conserved time evolution of such a system is described
by the normalized density operator
$\hat\rho$,
which
can be cast
in the form
\be\lb{eq:rhonorm}
\hat\rho = \densnnorm / {\rm tr}\, \densnnorm
,
\ee
where $\densnnorm$ is called the non-normalized
density operator.
This operator is defined as a solution
of the evolution equation
\be\label{e:Omega-dyna}
\frac{d}{dt}
\densnnorm
=-\frac{i}{\hbar}\left[\hat{H}_+,\densnnorm\right]
-\frac{1}{\hbar}\left\{\hat{\Gamma},\densnnorm\right\}
,
\ee
where the square brackets 
denote the commutator and the curly ones denote the anti-commutator.
This evolution equation can be directly derived from the \schrod equation,
see, for instance, ref. \cite{sz13}.
Further, in this equation one can change from $\densnnorm$ to $\hat\rho$, 
and obtain the evolution equation for
the normalized density operator itself
\begin{eqnarray}
\frac{d}{dt}
\hat\rho
=
-\frac{i}{\hbar}
\left[\hat{H}_+,\hat\rho
\right]
-
\frac{1}{\hbar}
\left\{\hat{\Gamma},\hat\rho 
\right\}
+\frac{2}{\hbar} 
\langle \Gamma \rangle
\hat\rho
,
\label{e:eomrho}
\end{eqnarray}
where we imply the standard definition for mean values,
$
\langle A \rangle = {\rm tr} (\hat\rho \, \hat{A})
$.

From the mathematical point of view, equations (\ref{e:Omega-dyna}) or (\ref{e:eomrho}), 
together with the definition for computing mean values,
represent the map that allows to describe the time evolution of  
system (\ref{e:nhham})
in terms of the matrix differential equation which is 
defined on similar axiomatic foundations as the conventional 
master equations of the Liouville and Lindblad kind  \cite{lin76,lsa06}.
According to this map, the Hermitian operator 
$\hat H_+ =  (\hat H + \hat H^\dagger)/2$ takes
over a role of the system's Hamiltonian (cf. the commutator in equations (\ref{e:Omega-dyna}) or (\ref{e:eomrho}))
whereas
the Hermitian operator $\hat\Gamma = i (\hat H - \hat H^\dagger)/2 $ induces
the additional terms in the evolution equation
that are supposed to account for new effects.
In other words, a theory with the non-Hermitian Hamiltonian $\hat H$ is dual to
a Liouvillian-type theory with the Hermitian Hamiltonian $(\hat H + \hat H^\dagger)/2$
but with the modified
evolution equation, which thus becomes the master equation of a special
kind.
This mapping not only 
reveals new features of the dynamics driven by non-Hermitian Hamiltonians 
but also facilitates their application 
for open quantum systems \cite{sz14}.

From the viewpoint of theory of open quantum systems,
the evolution equation
for the non-normalized density operator $\densnnorm$
effectively describes the original subsystem (with Hamiltonian $\hat{H}_+$) and the effect of 
environment (represented by $\hat{\Gamma}$).
Consequently,
the evolution equation
for the normalized density operator $\hat\rho$
effectively describes the original subsystem $\hat{H}_+$ together with the effect of 
environment  $\hat{\Gamma}$ 
and the probability flow between the subsystem and reservoir.
From the viewpoint of the subsystem alone,
this flow is an essentially non-Hamiltonian process since it is described not by means
of any kind of Hamiltonian but 
through the last term in the evolution equation (\ref{e:eomrho}).
This makes NH models somewhat similar to the Lindblad-type ones,
where the effect of the environment is encoded in the evolution equation
through the additional term often dubbed the dissipator.
However, the important difference is that the Lindblad dissipator, which is a traceless
operator by construction, does not affect the conservation of probability of the system
whereas the last term in (\ref{e:eomrho}) restores the probability's conservation
which is otherwise broken by the anticommutator term.
Besides, one can see that the last term in (\ref{e:eomrho})
is nonlinear with respect to $\hat\rho$, unlike its Lindblad analogue.
It is interesting that the appearance of nonlinearities in NH-related theories has
been also suggested some time ago, although on different grounds of the Feshbach-Fano projection
formalism \cite{zno02}.

Further, if one introduces the quantum purity ${\cal P} = \text{tr} (\hat\rho^2) $
then one can show that
its time evolution
is governed by the
equation
\be\lb{e:purder}
\frac{d}{d t}
{\cal P}
=
{\cal R} (\hat\rho)
\equiv
\frac{4}{\hbar}
\left[
\langle \Gamma \rangle
{\cal P}
-
\text{tr} (\hat\rho^2 \,\hat\Gamma)
\right]
,
\ee
where ${\cal R} (\hat\rho)$ is the purity rate function \cite{sz13}.
It is easy to see that 
the rate function vanishes identically
in the case of Hermitian evolution ($\hat\Gamma = 0$),
but is otherwise an
essentially non-trivial function of both the density
operator and the anti-Hermitian part of Hamiltonian.
As long as the density matrix $\hat\rho_p$ of a pure state, which is
defined via 
the idempotence $\hat\rho_p^2 = \hat\rho_p$,
is the equilibrium point in both the Hermitian and
non-Hermitian cases (\textit{i.e.}, ${\cal R} (\hat\rho_p) = 0$),
one might expect that any pure state is always preserved
during the NH evolution.

However, this could only be possible if one disregards the
other important player in the quantum realm -- quantum fluctuations.
Indeed, if an (initially) pure state is not protected against the fluctuations that
can alter its purity, then during time evolution it will be driven away from being
pure, no matter how small these fluctuations initially were.
This phenomenon of dynamical instability is not directly seen in 
the evolution equation for the density operator, but emerges via solutions thereof.
It is somewhat analogous to the spontaneous symmetry breaking
in field theory -- except that here one deals not with the actual
field potential, 
but with some kind of the fictitious-particle potential function 
(usually dubbed as the Lyapunov function candidate)
that determines whether the equilibrium point $\hat\rho_p$ is stable or not.
This also makes NH models different from the Lindblad-type ones,
in which the non-conservation of purity and pure states directly follows from 
the underlying evolution equation.
Besides, the Lindblad master equation approach has a different range of applicability
because it implies a number of certain approximations, such as
the Markovian, Born and rotating-wave ones, whereas NH Hamiltonians often appear  
in theories in a more direct way, 
one example to be the Feshbach-Fano projections \cite{fesh58}.

It is worth  mentioning also that the above-mentioned kind of chaotic behaviour should
not be confused with the notion of ``quantum chaos'', which is currently 
being reserved in the field of research devoted to how classical chaotic dynamical systems 
can be described by means of methods and concepts of quantum mechanics \cite{chi79,zas81,ber03}.

\scn{General stability approach}{s-gen}

In order to develop the general approach for performing the stability analysis of pure states, 
let us introduce the non-purity (mixedness) operator
\be
\mixmat = \hat\rho - \hat\rho^2
,
\ee
whose trace is known
as the linear entropy \cite{zur93} 
\be
S_L = \text{tr} \mixmat  = 1 - {\cal P}
.
\ee
In order to consider variations of the density operator
around some pure state $\hat\rho_p$,
we perform the decomposition
\be\lb{e:rhodecomp}
\hat\rho = \hat\rho_p + \drho
, 
\ee
where $\drho$ is the variation operator.
As long as the main properties of the
density matrix 
should be left intact by such decomposition,
$\drho$ must be Hermitian and traceless;
this automatically ensures the well-defined probability
and real mean values of operators.
One obtains that
\be
\hat\rho^2 = \hat\rho_p + \left\{ \hat\rho_p, \drho \right\} 
+ \drho^2
,\ \
\mixmat 
= 
\drho
-
\left\{ \hat\rho_p, \drho \right\} 
- \drho^2
,
\ee
such that the operator $\mixmat$
is clearly a measure
of deviation of a state from being pure.
Consequently, equations (\ref{e:eomrho}) and (\ref{e:purder})
yield the equations
that are more suitable for stability analysis,
\bw
\ba
\frac{d}{dt}\drho
&=&
-\frac{i}{\hbar}
\left[\hat{H}_+, \drho
\right]
-
\frac{1}{\hbar}
\left\{\hat{\Gamma}, \drho 
\right\}
+
\frac{2}{\hbar} \hat\rho_p 
\dgammatr 
+\frac{2}{\hbar} 
\left(
\langle \Gamma \rangle_p
+
\dgammatr
\right)
\drho
,
\label{e:eomdrho}\\
\frac{d}{d t}
\mixmat
&=&
-\frac{i}{\hbar}
\left[\hat{H}_+, \mixmat
\right]
-
\frac{1}{\hbar}
\left\{\hat{\Gamma}, \mixmat 
\right\}
+\frac{4}{\hbar} 
\left(
\langle \Gamma \rangle_p
+
\dgammatr
\right)
\mixmat
+
\frac{2}{\hbar} 
\left[
(\hat\rho_p + \drho) \drho^{(\Gamma)}
+
\drho \hat\rho_p^{(\Gamma)}
\right]
,
\lb{e:eompurmat}\\
\frac{d}{d t}
S_L
&=&
\frac{4}{\hbar} 
\left[
\left(
\langle \Gamma \rangle_p
+
\dgammatr
\right)
S_L
-
\text{tr} (\hat\Gamma \mixmat)
\right]
,
\lb{e:eomdentr}
\ea
\ew
where 
$\langle \Gamma \rangle_p = {\rm tr} (\hat\rho_p \, \hat{\Gamma}) $
is the average of the operator $\Gamma$ with respect to the
unperturbed density operator $\hat\rho_p$,
$\dgammatr = {\rm tr} (\drho \hat{\Gamma}) $
is the difference between the averages
$\langle \Gamma \rangle$ and $\langle \Gamma \rangle_p$,
and 
we have used the notation 
$\hat A^{(\Gamma)} = \hat\Gamma \hat A - \text{tr} (\hat\Gamma \hat A) \hat I$
with $\hat I$ being the identity operator.
One can easily verify that both the tracelessness and
hermiticity of the variation operator $\drho$ are preserved during time evolution.

Equations (\ref{e:eomdrho})-(\ref{e:eomdentr})  
are matrix differential equations that
can all be used for the stability study, but
it must be emphasized that equations (\ref{e:eompurmat}) and (\ref{e:eomdentr}) 
describe only those variations that can potentially lead to the transition
of a pure state $\hat\rho_p$ into a mixed one (dubbed as the ``mixing'' fluctuations
in what follows),
whereas equation (\ref{e:eomdrho}) alone
governs the variations of a general type, regardless on whether they alter the
purity of $\hat\rho_p$ or not.
Therefore, the equations (\ref{e:eompurmat}) and (\ref{e:eomdentr}) will be
of special interest here.
It is easy to see that one can apply to them the standard
methods of dynamical stability analysis, such as the Lyapunov
or Vakhitov-Kolokolov criteria \cite{cbook}.

However, some generic features can be noticed straight away.
In particular, equation (\ref{e:eomdentr}) reveals that stability 
against the ``mixing'' fluctuations essentially depends
on the result of competition between the terms
$\left\langle \Gamma \right\rangle $ and 
$\text{tr} (\hat\Gamma \mixmat) / S_L = 
(
\left\langle \Gamma \right\rangle - \text{tr} (\hat\rho^2 \hat\Gamma)
)/S_L$.
If their difference is non-negative at any time, then during evolution the fluctuations will
keep the initial state away from being pure. 
If the difference is negative then the  fluctuations will get suppressed. 
 
Yet another property, which can be immediately seen from equation (\ref{e:eomdentr}), is that,
in the absence of the anti-Hermitian part of the Hamiltonian, random fluctuations
of a given pure state would never get suppressed but their magnitude remains at much the same level.
This means that in order to 
fully suppress 
fluctuations,
suitably chosen anti-Hermitian terms must be present to the Hamiltonian.
In those cases the anti-Hermitian part would ensure 
the full stability of pure states even if it is negligibly small compared
to the Hermitian part.

\scn{Local stability}{s-loc}

It is clear that the local instability of a pure state against fluctuations
leading to the mixing of a state does yet not imply 
global instability (\textit{i. e.}, when the purity of a perturbed pure state never 
goes back to its original value $1$).
However, this case is still interesting from 
a physical point of view: quantum systems might exist  in which 
the purification time (\textit{i. e.}, the time of return back into a pure
state) can be larger than the lifetime of the system itself
or the ultimately possible time of measurement/observation of the system.
Yet another application area of local instability is that
it can point to the presence of singular points at which
the density matrix components diverge; thus indicating that
the underlying system becomes critically unstable.
In any of these cases, local instability of a pure state might become 
the dominating chaotic-type phenomenon, which can
strongly affect the physical properties of systems' evolution.
From the technical point of view, the advantage of the study of local instability
is that one does not have to know the whole solution for the density operator,
but only a few characteristic exponents.

The analysis of local stability for a given state of the NH quantum system
can be made based on linearized equations, which are easier to solve
or analyze using well-known methods from the theory of stability and
dynamical chaos.
Indeed, once we assume that the variation operator 
is small, $\drho = \delta\hat\rho$, we can perform the linearization with respect
to it. 
Then equations (\ref{e:eomdrho})-(\ref{e:eomdentr}) yield,
respectively:
\bw
\ba
\frac{d}{dt}\frho
&=&
-\frac{i}{\hbar}
\left[\hat{H}_+, \frho
\right]
-
\frac{1}{\hbar}
\left\{\hat{\Gamma}, \frho 
\right\}
+\frac{2}{\hbar}
\left[ 
\langle \Gamma \rangle_p \frho
+
\hat\rho_p {\rm tr}
(\frho \, \hat{\Gamma})
\right]
+ {\cal O} (\frho^2)
,
\label{e:eomfrho}\\
\frac{d}{d t}
\delta \mixmat
&=&
-\frac{i}{\hbar}
\left[\hat{H}_+, \delta \mixmat
\right]
-
\frac{1}{\hbar}
\left\{\hat{\Gamma}, \delta \mixmat 
\right\}
+\frac{2}{\hbar}
\left[
2 
\langle \Gamma \rangle_p
\delta \mixmat
+
\hat\rho_p  \drho^{(\Gamma)}
+
\hat\rho_p^{(\Gamma)} \frho
\right]
+ {\cal O} (\frho^2)
,
\lb{e:eomfpurmat}\\
\frac{d}{d t}
\delta S_L
&=&
\frac{4}{\hbar} 
\left[
\langle \Gamma \rangle_p \,\delta S_L
-
\text{tr} (\hat\Gamma \delta\mixmat)
\right]
+ {\cal O} (\frho^2)
,
\lb{e:eomfpur}
\ea
\ew
where $\delta\mixmat = \mixmat |_{\drho \to \frho} = \frho - \left\{ \hat\rho_p, \frho \right\} $
and $\delta S_L = \text{tr} (\delta\mixmat ) = - 2 \text{tr} (\hat\rho_p \, \frho) = - \fpur$
are the variations of the non-purity operator and its trace, respectively;
${\cal O}$ denotes the terms that can be neglected in the linear order of approximation.

As discussed above, in order to study the stability of a system
against the ``mixing'' fluctuations,
one has to consider primarily equations (\ref{e:eomfpurmat}) and (\ref{e:eomfpur}).
In the leading approximation (\ref{e:eomfpurmat}) can be rewritten in the form 
$
\frac{d}{dt} \charvec
=
\charmat \charvec
$,
where
$\charvec$ is a column vector constructed out of the linearly independent 
components of the matrix 
$\delta \mixmat$,
and $ \charmat = \charmat (\hat\rho_p, \hat H_+, \hat\Gamma, t)$
is a characteristic matrix.
If all real parts of its eigenvalues are negative then the pure state 
$\hat\rho_p$ is locally stable against small ``mixing'' fluctuations.
This condition is equivalent to the matrix $\charmat\!^T  \pdmat + \pdmat \charmat$
being negative definite for some positive definite matrix $\pdmat = \pdmat^T$.
The corresponding Lyapunov function candidate can be determined then as
$V (\charvec) = (\charvec)^T \pdmat\, \charvec $.
Further classification of instability types 
(``center'', ``node'', ``saddle'', ``spiral'', \textit{etc.}) 
can be done for a specific system
by analysis of the characteristic matrix's eigenvalues \cite{cbook}.

\scn{Example: two-level systems}{s-ex}

In order to illustrate the formalism, let us consider some models 
which exhibit, under certain conditions, the dynamical instability of pure
states.

The NH two-level systems (TLS) are simple yet very
useful physical models \cite{datto}, which can provide 
a clear visualization of different kinds of stability that may occur. 
One can check that for two-level systems 
the non-purity matrix has only one independent component
and can be written in the form 
\be
\mixmat = \frac{1}{2} S_L \hat I
,
\ee
then one obtains
\be
\delta\mixmat = \frac{1}{2} \delta S_L \hat I
,
\ee
\textit{i. e.}, the scalar
$\delta S_L$ provides complete information
about the local stability of a given pure state $\hat\rho_p$,
whereas the characteristic matrix $\charmat$ can be reduced to
a $1\times 1$ matrix.
The linearized equation (\ref{e:eomfpur}) takes the simple
form
\be
\frac{1}{\delta S_L}
\frac{d}{d t}
\delta S_L
=
\Lambda (\hat\rho_p, \, \hat\Gamma )
\equiv
\frac{2}{\hbar} 
\left(
2
\langle \Gamma \rangle_p 
-
\text{tr}\,\hat\Gamma
\right)
,
\lb{e:eomfpurTLS}
\ee
where $\Lambda = \Lambda (\hat\rho_p, \, \hat\Gamma )$ is the  characteristic exponent.
According to the approach,
the state $\hat\rho_p$ is locally stable 
against the ``mixing'' fluctuations if  
$\Lambda (\hat\rho_p, \, \hat\Gamma ) < 0$, and locally unstable otherwise. 
To illustrate this, let us consider the following exactly-solvable models.


\sscn{Tunneling models with non-Hermitian detuning}{s-tls-1}

This is a set of NH models whose Hermitian and anti-Hermitian parts of 
Hamiltonian are, respectively:
\be\lb{e:hamtls1}
\hat H_+ = - \hbar \rabi \hat\sigma_x,
\ \
\hat\Gamma = \hbar \lambda \hat\sigma_z
,
\ee
where $\hat\sigma$'s are Pauli matrices,
positive
parameter $\rabi$ is related to the matrix element for tunneling between two wells, 
and $\lambda$ is a constant parameter, real-valued but otherwise free, that
can be viewed as
the imaginary counterpart of the detuning parameter \cite{leg87}.
Such models are popular in many areas of quantum physics, including 
the theory of open quantum-optical systems 
(in particular, when describing
the direct photodetection of a driven two-level atom
interacting with the electromagnetic field -- in which
case $\lambda$ would be related to the atomic damping rate)
\cite{bpbook,sz14}
and non-Hermitian quantum mechanics 
with real energy spectra \cite{sgh92,ben98}.

Here we would like to determine for which values of the parameters $\rabi$
and $\lambda$ a pure state, say, the one defined by
\be\lb{e:rhopurM1}
\hat\rho_p^{(1)} = 
|e\rangle\langle e|  
= \twomat 1 0 0 0
,
\ee
is locally stable against the ``mixing'' fluctuations.
In case of quantum-optical or spin systems, this information could be instrumental for
determining whether a system can undergo, respectively, the spontaneous emission
or spin-flip transition.

The characteristic exponent $\Lambda$, which can be 
immediately computed from equation  (\ref{e:eomfpurTLS}),
turns out to be equal to $\lambda$,
up to a positive factor.
Therefore, we expect the state $\hat\rho_p^{(1)}$ to be locally stable against
small fluctuations for the models with negative $\tilde\lambda = \lambda / \rabi$,
and locally unstable otherwise.

To verify this, one needs to solve the evolution equation
assuming the perturbed initial conditions:
\be\lb{e:rhoiniptb}
\hat\rho (0) = \hat\rho_p^{(1)} + \delta\hat\rho_{(0)}
,
\ee
where, according to the approach, the variation matrix can be chosen
in the form
\be\lb{e:varmattls}
\hat\Delta_{(0)} = \delta\hat\rho_{(0)} =
\twomat{\delta_2}{\delta_1}{\delta_1}{-\delta_2}
,
\ee
with $\delta_i$'s being arbitrary real-valued numbers 
(for simplicity 
we have omitted the off-diagonal imaginary components of $\hat\Delta$).
The positivity of 
the total density operator $\hat\rho (0)$ can be always ensured by imposing constraints 
on the values of variation matrix's components.
However, here we assume the fluctuations to be arbitrary enough in a sense 
that we do not postulate them to exactly preserve the positivity 
of the density operator at the initial moment of time, 
similarly to the approach \cite{sz14cor}. 
Instead, one might be interested to see whether this property
can be dynamically restored
during evolution if it has been initially broken by fluctuations. 

Solving the evolution equation (\ref{e:eomrho})
with the initial condition (\ref{e:rhoiniptb}),
we obtain the following expression for the normalized density operator:
\be
\hat\rho  (t)
=
\frac{f_x (t)}{F(t)} \hat\sigma_x
+
\frac{f_y (t)}{F(t)} \hat\sigma_y
+
\frac{f_z (t) }{2 F(t)}
\hat\sigma_z
+ 
\frac{1}{2}
\hat I
,
\ee
where
we have denoted:
\ba
&&
f_x (t) =
\delta_1 \mu^2
,\\&&
f_y (t) =
\sinh{(\mu \tau)}
\left[
\mu  p_2  \cosh{(\mu \tau)}
-
\tilde\lambda \sinh{(\mu \tau)}
\right]\!,~~~
\\&&
f_z (t) =
\mu
\left[
\mu  p_2  \cosh{(2 \mu \tau)}
-
\tilde\lambda \sinh{(2 \mu \tau)}
\right]
,
\\&&
F (t) =
\tilde\lambda^2 \cosh{(2 \mu \tau)}
- 
p_2
\mu \tilde\lambda \sinh{(2 \mu \tau)}
-
1
,
\ea
where
$\tau = \omega t$, $\tilde\lambda = \lambda/ \omega$,
$p_2 = 2 \delta_2 + 1$
and $\mu = \sqrt{\tilde\lambda^2 -1}$.

\begin{figure}
\centering
\subfloat[$\tilde\lambda = 1/2$]{
  \includegraphics[width=0.49\columnwidth]{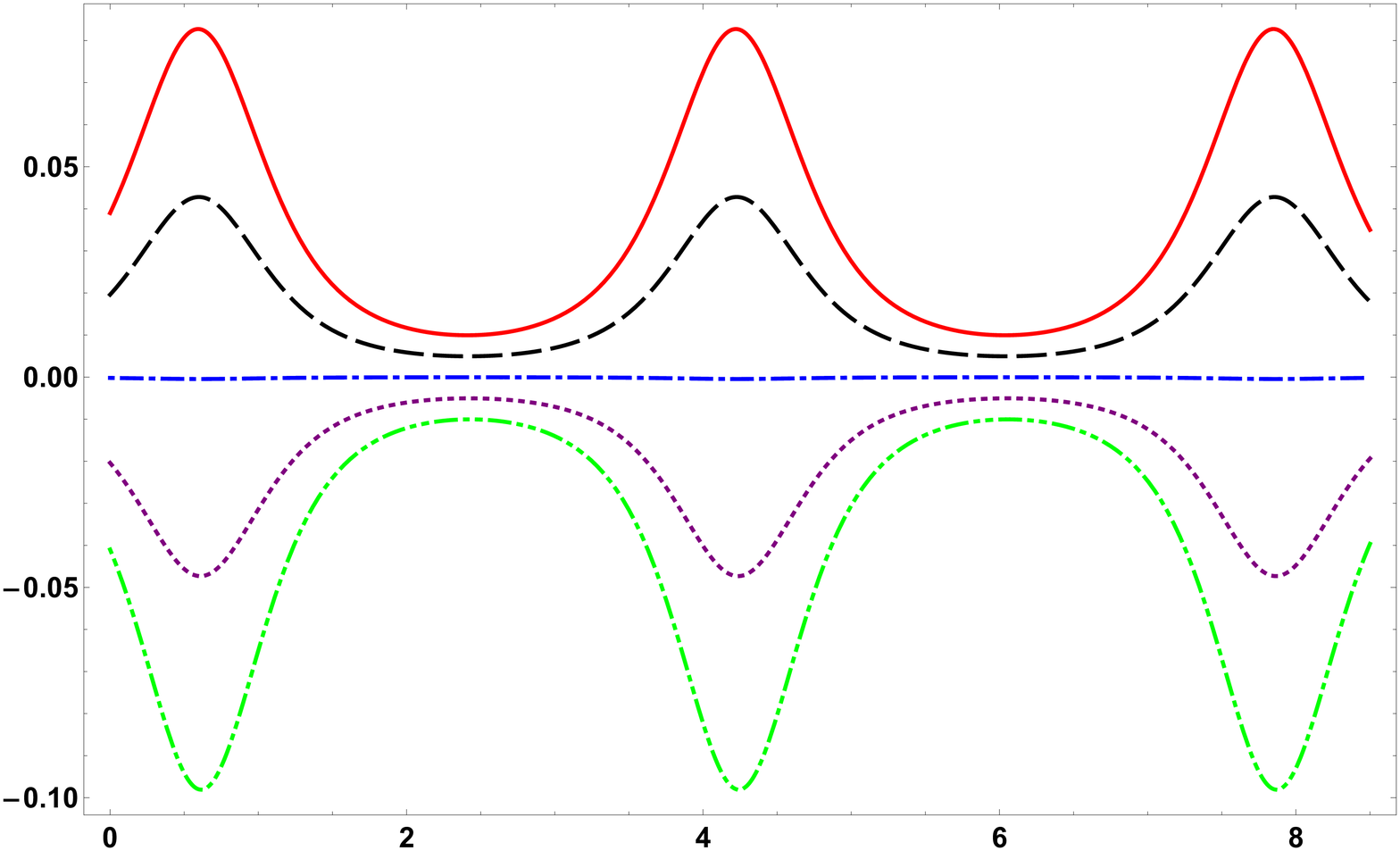}
}
\subfloat[$\tilde\lambda = 2$]{
  \includegraphics[width=0.49\columnwidth]{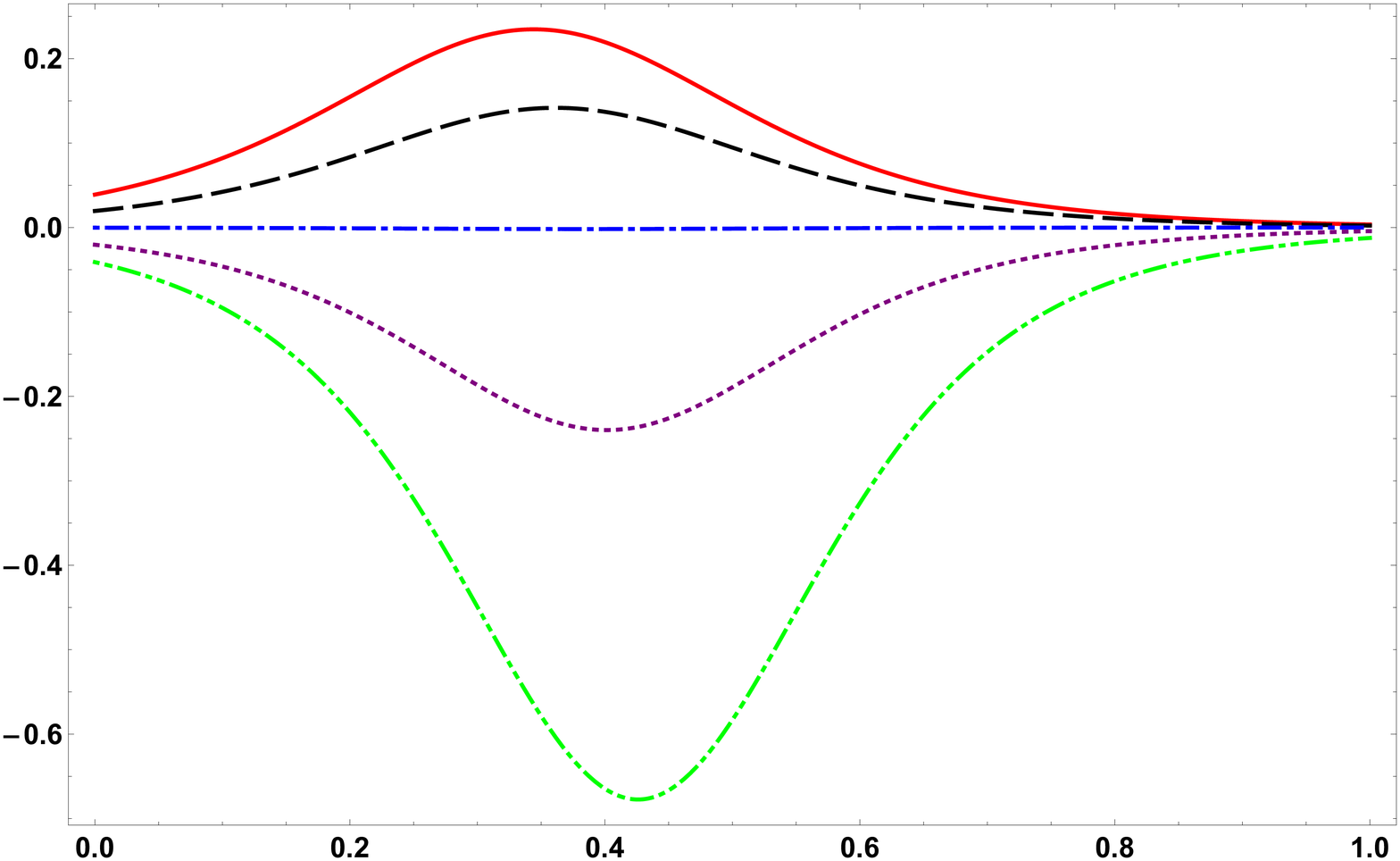}
}
\hspace{0mm}
\subfloat[$\tilde\lambda = -1/2$]{
  \includegraphics[width=0.49\columnwidth]{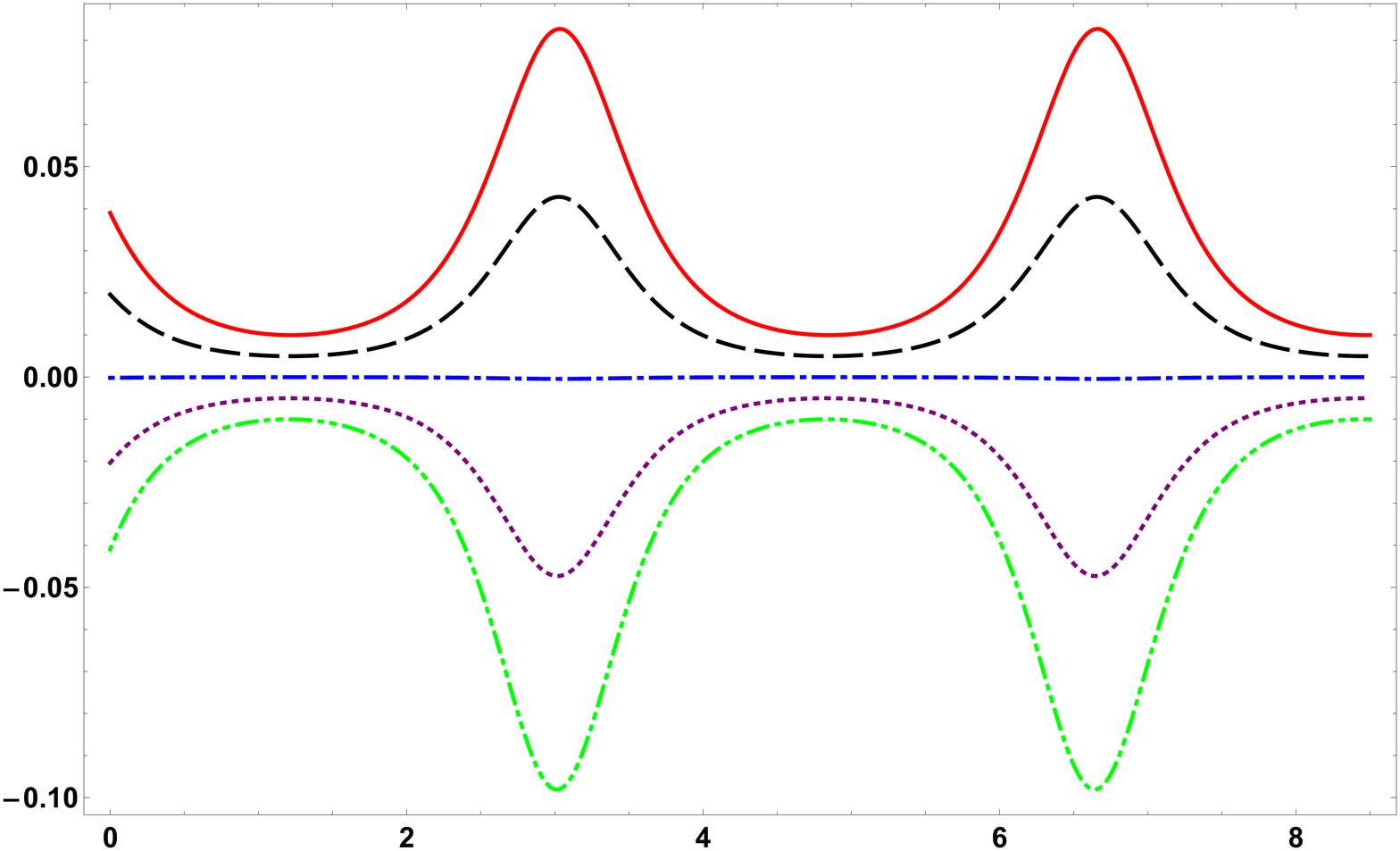}
}
\subfloat[$\tilde\lambda = -2$]{
  \includegraphics[width=0.49\columnwidth]{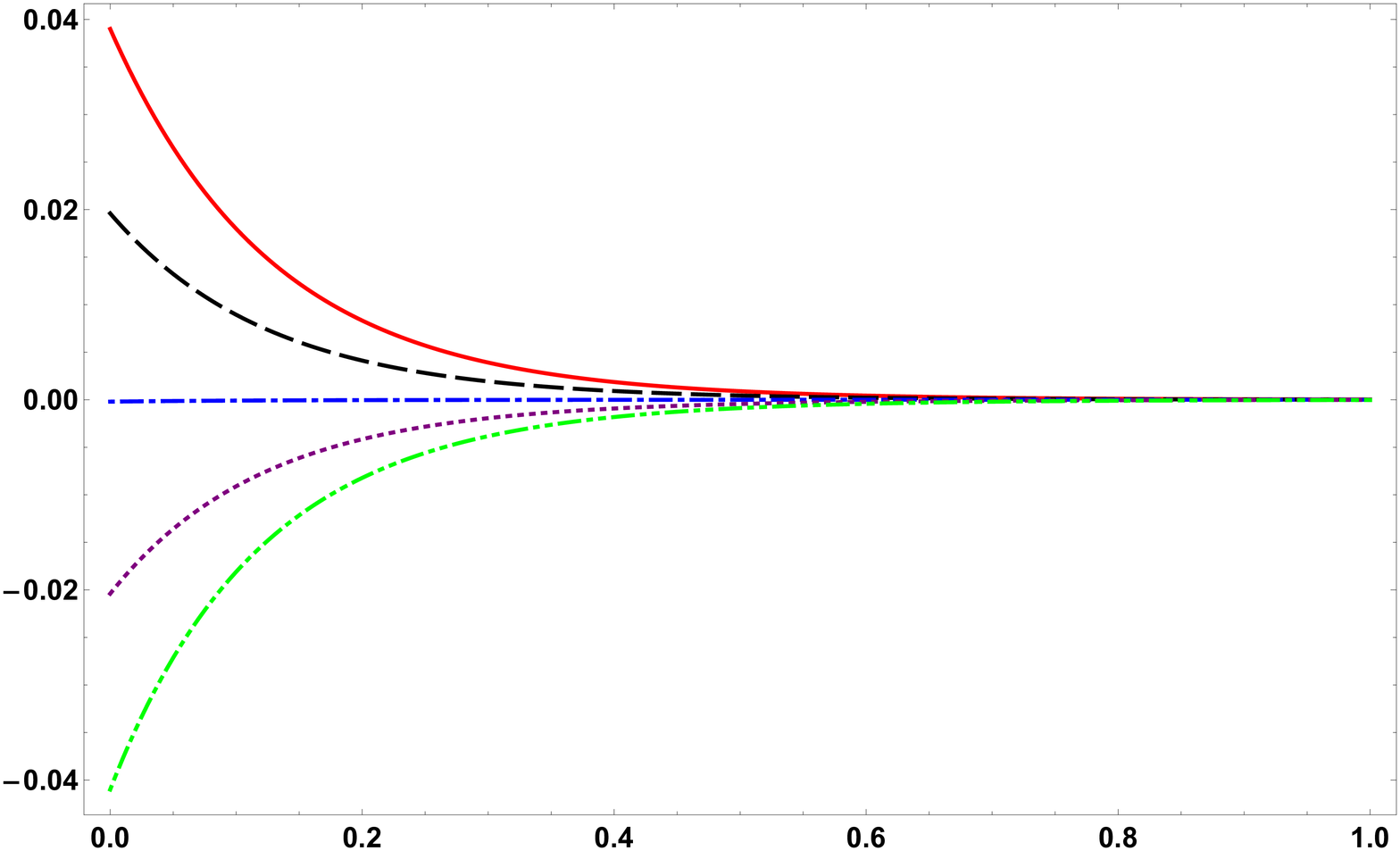}
}
\caption{Linear entropy $S_L$ 
versus time (in units of $\rabi^{-1}$) for
the family of models (\ref{e:hamtls1}), at different
values of $\tilde\lambda$ and initial perturbations of the state $\hat\rho_p^{(1)}$. 
The value of $\delta_1$ is $0.01$ for all curves, 
the values of $\delta_2$ are: 
$-0.02$ (solid curves),
$-0.01$ (dashed curves), 
$0$ (dash-dotted curves), 
$0.01$ (dotted curves) 
and 
$0.02$ (dash-double-dotted curves).
}
\label{f-mod1}
\end{figure}

The evolution of purity for this solution 
is presented in Fig. \ref{f-mod1}.
The top (bottom) row of the figure 
shows models for which the initial state $\hat\rho_p^{(1)}$ is unstable (stable) locally,
as predicted by theory.
Indeed, one can see that for positive values of $\tilde\lambda$, 
which has the same sign as $\lambda$,
the magnitude  of the linear entropy initially increases from its initial value,
whereas for negative $\tilde\lambda$'s the magnitude  of the linear entropy
decreases straight from the beginning. 
We reiterate here that the local stability or instability does  yet not imply
the global one.
However, if the measurement time or the lifetime of a system is limited and smaller
than the purification time (which is of the scale $\rabi^{-1}$ here), 
then an observer would not be able to empirically distinguish between the
local and global (asymptotical) types of instability.  

Further, the left (right) column of Fig. \ref{f-mod1}
shows models for which the initial state $\hat\rho_p^{(1)}$ is unstable (stable) asymptotically.
Asymptotical stability cannot be assessed based on 
the linearized equation (\ref{e:eomfpurTLS}) but
requires the usage of a general approach (\ref{e:eomdrho})-(\ref{e:eomdentr}),
which is quite bulky and thus would bring us beyond our purposes here.
Instead, in Figs. \ref{f-mod1}b and \ref{f-mod1}d we demonstrate that 
corresponding models
contain the mechanism that asymptotically suppresses quantum fluctuations,
whereas Figs. \ref{f-mod1}a and \ref{f-mod1}c reveal that 
in some models
fluctuations may keep the state away from being pure in an oscillatory way.
As long as both the conventional von Neumann entropy 
and the NH-adapted entropy 
(defined in Ref. \cite{sz15}) would also oscillate in those cases, one can regard this behaviour as the quantum non-Hermitian analogue of the self-organization phenomenon.


\sscn{Tunneling models with analytically continued matrix element}{s-tls-2}

This is a family of models where
\be\lb{e:hamtls2}
\hat H_+ = - \hbar \rabi \hat\sigma_x,
\ \
\hat\Gamma = \hbar \eta \hat\sigma_x
,
\ee
with
$\eta$ being a constant parameter, real-valued but otherwise free.
Such models describe physical cases when a quantum environment of some kind
(external fields or vacuum oscillations) effectively shifts the value of
the tunneling parameter into a complex domain: 
$\rabi \to \rabi + i\,\eta$.

Here we would like to determine for which values of the parameters 
$\rabi$ and $\eta$ 
the pure state
\be\lb{e:rhopurM2}
\hat\rho_p^{(2)} 
= \frac{1}{2} \begin{pmatrix} 1 & 1 \\ 1 & 1 \end{pmatrix}
\ee
is locally stable against the ``mixing'' fluctuations.
From the physical point of view, it might be used to
determine whether a given quantum superposition of states is protected
against the spontaneous decay into a mixed state.

The characteristic exponent (\ref{e:eomfpurTLS}),
turns out to be equal to $\eta$,
up to a positive factor, hence, we expect state $\hat\rho_p^{(2)}$ to be locally stable against
small fluctuations in the models with negative $\tilde\eta = \eta / \rabi$,
and locally unstable otherwise.

To verify this, one needs to solve the evolution equation
(\ref{e:eomrho})
assuming the perturbed initial conditions
\be\lb{e:rhoiniptb2}
\hat\rho (0) = \hat\rho_p^{(2)} + \delta\hat\rho_{(0)} 
,
\ee
where $\delta\hat\rho_{(0)} $ is given by (\ref{e:varmattls}). 
We obtain the following expression for the normalized density operator:
\be
\hat\rho  (t)
=
\frac{g_x (t)}{2 G(t)} \hat\sigma_x
+
\frac{g_y (t)}{G(t)} \hat\sigma_y
+
\frac{g_z (t) }{G(t)}
\hat\sigma_z
+ 
\frac{1}{2}
\hat I
,
\ee
where
we have denoted:
\ba
&&
g_x (t) =
p_1
\cosh{(2 \tilde\eta \tau)}
- 
\sinh{(2 \tilde\eta \tau)}
,\\&&
g_y (t) 
=
\delta_2
\sin{(2 \tau)}
,
\\&&
g_z (t) =
\delta_2
\cos{(2 \tau)}
,
\\&&
G (t) =
\cosh{(2 \tilde\eta \tau)}
- 
p_1 \sinh{(2 \tilde\eta \tau)}
,
\ea
where
$\tau = \omega t$, $\tilde\eta = \eta/ \omega$,
and
$p_1 = 2 \delta_1 + 1$.

\begin{figure}
\centering
\subfloat[$\tilde\eta = -2$]{
  \includegraphics[width=0.49\columnwidth]{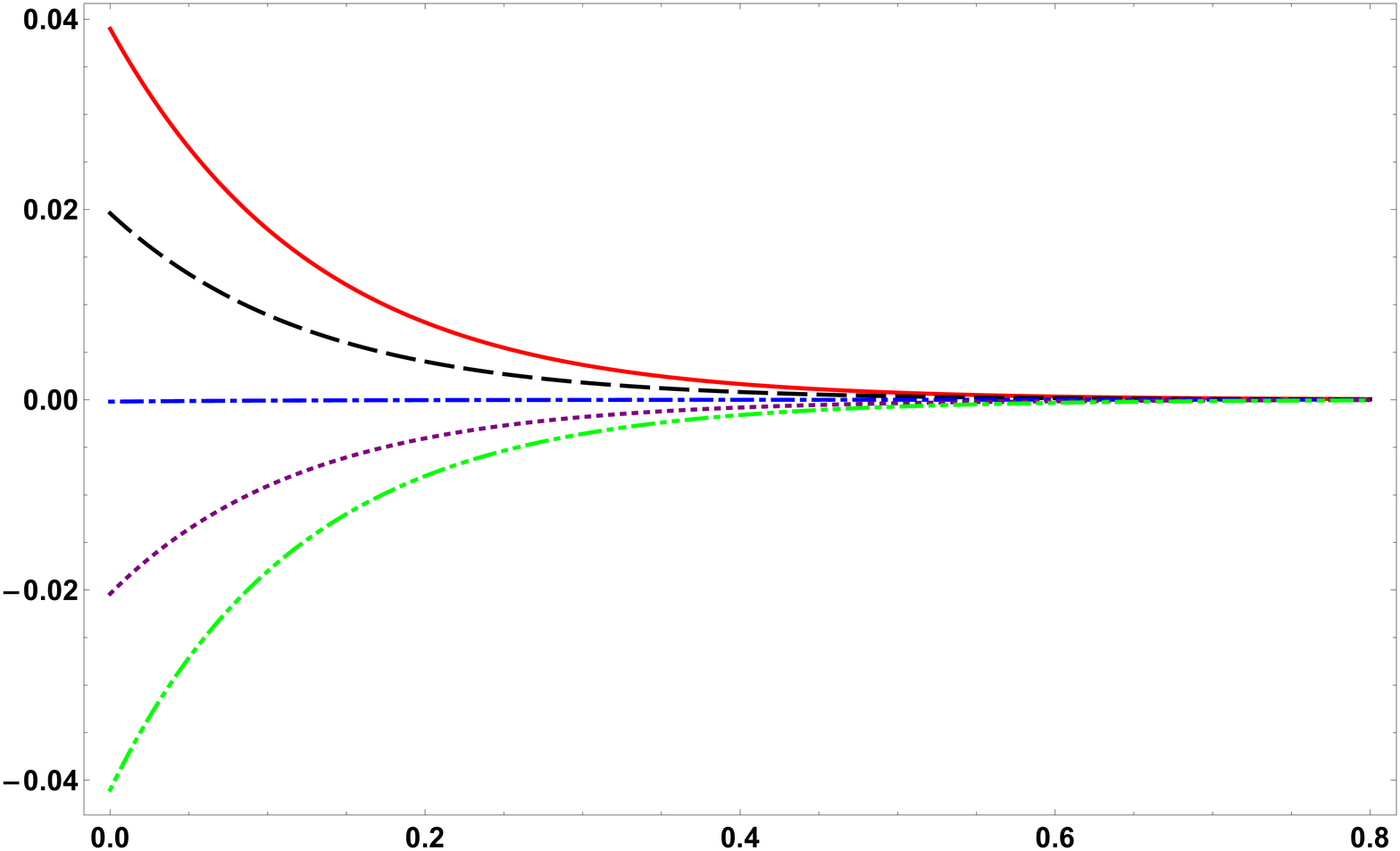}
}
\subfloat[$\tilde\eta = 2$]{
  \includegraphics[width=0.49\columnwidth]{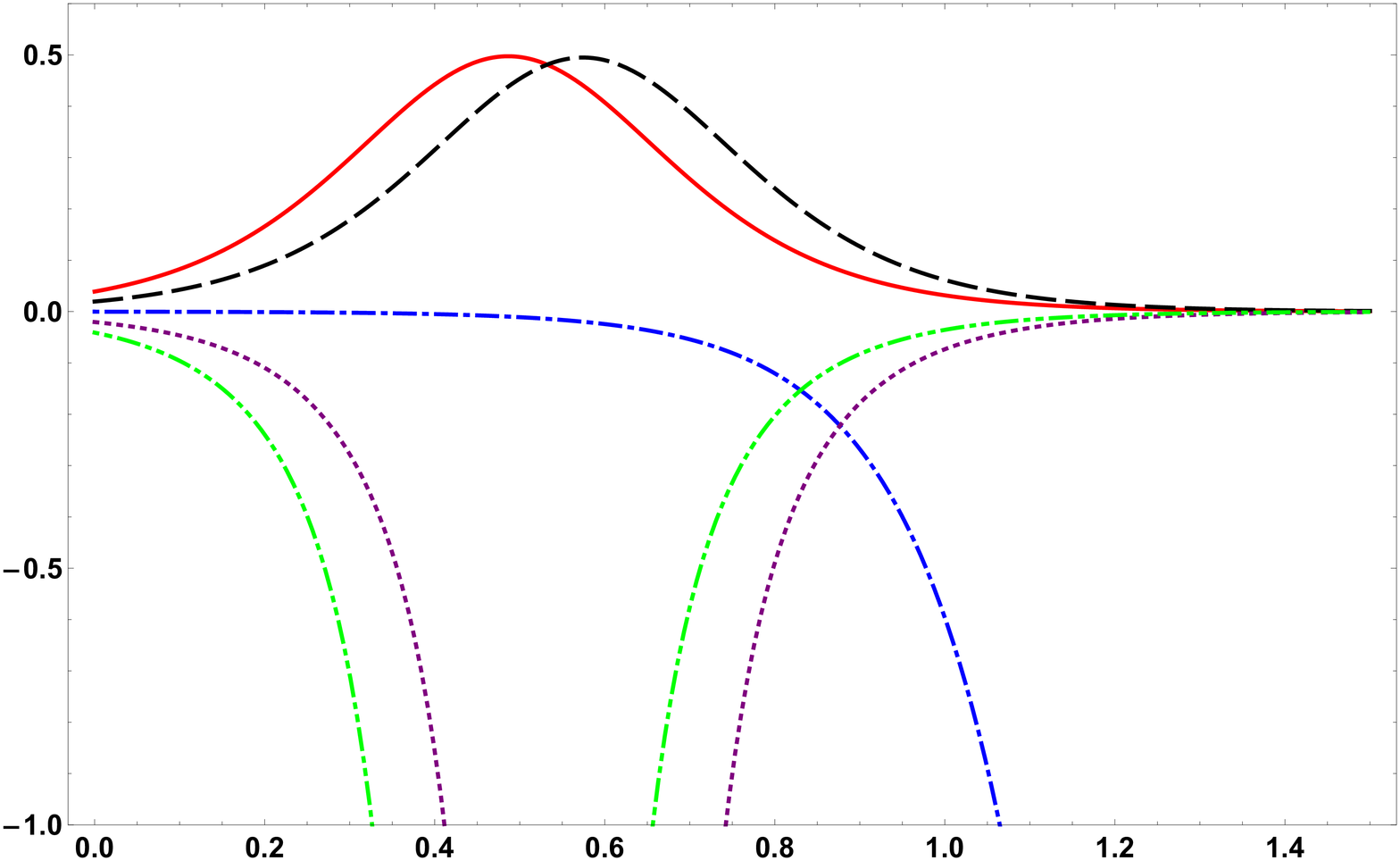}
}
\hspace{0mm}
\caption{Linear entropy  $S_L$ 
versus time (in units of $\rabi^{-1}$) for
the family of models  (\ref{e:hamtls2}), at different
values of $\tilde\eta$ and initial perturbations of the state $\hat\rho_p^{(2)}$. 
The value of $\delta_2$ is $0.01$ for all curves, 
the values of $\delta_1$ are: 
$-0.02$ (solid curves),
$-0.01$ (dashed curves), 
$0$ (dash-dotted curves), 
$0.01$ (dotted curves) 
and 
$0.02$ (dash-double-dotted curves).
}
\label{f-mod2}
\end{figure}

The evolution of purity for this solution 
is presented in Fig. \ref{f-mod2}.
Figure \ref{f-mod2}a illustrates the model for which the initial state $\hat\rho_p^{(2)}$ 
is stable locally, as predicted by the linearized approach. 
It also turns out to be stable  globally,
which makes this case similar to the one discussed earlier, regarding Fig. \ref{f-mod1}d.

In Fig. \ref{f-mod2}b we show the model for which some  
fluctuation modes of the initial state $\hat\rho_p^{(2)}$ grow indefinitely,
so that the system evolves into a singularity.
This is another kind of dynamical instability, which is different from
those discussed in the previous set of models.
It illustrates that for some models, 
quantum fluctuations either get significantly amplified during
certain interval of time (solid and dashed curves in Fig. \ref{f-mod2}b) 
or destabilize the original state
up to the full destruction
(dotted, dash-dotted and dash-double-dotted curves).

\scn{Conclusion}{s-con}

Using the density operator approach for
non-Hermitian Hamiltonians,
we have demonstrated that quantum fluctuations can cause, 
under certain conditions, the instability of pure states, which
is controlled by the environment-induced anti-Hermitian parts of Hamiltonians. 
This is drastically different from the Hermitian case where both the purity's
value and pure states are preserved during time evolution. 

It is shown that the instability of pure states
is not preassigned in the evolution equation but arises as the emergent
phenomenon in its solutions.
We have derived the equations that are necessary to study the
stability issues of any quantum system, regardless of the number
of its degrees of freedom, dimensionality of Hilbert space, etc.
Thus, the formalism and main results are applicable for systems described
by non-Hermitian Hamiltonians of general type.

Finally, in order to illustrate the different types of
instability that may occur, we have considered some exactly solvable two-state models.
Apart from being instructive examples on their own, these
models can be used in a theory of open quantum-optical and spin systems
where our stability analysis might be helpful in achieving
a better understanding of such phenomena as the spontaneous emission, decay or spin flip
(which, as a matter of fact, could be one of the directions for future work).
By means of those models we have visualized the main types of stability
of a pure state against small ``mixing'' fluctuations
that may occur in the open quantum system:
fluctuations get permanently suppressed with time (the state is locally and globally stable),
fluctuations amplify during a finite period of time
but eventually get suppressed (the state is locally unstable but asymptotically stable),
fluctuations never get suppressed with time but stay 
bound by a finite value (the state is globally unstable),
and, finally,
fluctuations indefinitely grow thus leading to the singularity 
and critical instability of the system (the state and system are globally unstable).
It should be also noticed that for the two-state models studied above the fluctuations, which 
cause the purity to acquire unphysical values, either get suppressed or lead to 
the overall instability of the system.

\begin{acknowledgments}
Discussions with D. C. Brody,  E.-M. Graefe and A. Sergi at the mini-workshop ``Quantum Dynamics and Non-Hermitian Hamiltonians'' (3-8 December 2014, Pietermartizburg, South Africa), 
during which the main ideas of this work have been reported, are acknowledged.
I also thank to A. Sergi for his support and hospitality during
my numerous visits to the University of KwaZulu-Natal in Pietermaritzburg.
The proofreading of the
manuscript by P. Stannard is greatly acknowledged as well.
This work is partially supported by the National Research Foundation of South Africa.

\end{acknowledgments}


\end{document}